\documentclass{elsart}

\usepackage{graphicx,xspace,epsfig,amssymb,bm}
\usepackage{dcolumn}

\newcommand{\cher}{Cherenkov\xspace}
\newcommand{\gsim}{\lower.7ex\hbox{$\;\stackrel{\textstyle>}{\sim}\;$}}
\newcommand{\lsim}{\lower.7ex\hbox{$\;\stackrel{\textstyle<}{\sim}\;$}}
\newcommand{\nuN}{$\nu N$}

\newcommand{\crss}{cross-sections}

\newcommand{\stfs}{structure functions}

\newcommand{\aniswww}{\mbox{http://www-zeuthen.desy.de/nuastro/anis/anis.html}}

\def\d{{\rm d}}

\begin{document}

\begin{frontmatter}
\title{ANIS: High Energy Neutrino Generator for Neutrino Telescopes}
\author{A.~Gazizov$^1$}
\author{and M.~Kowalski$^{1,2}$\corauthref{cor}}\corauth[cor]{Corresponding author. Electronic mail: marek.kowalski@desy.de}

\address{$^1$DESY Zeuthen, D-15735, Zeuthen, Germany}
\address{$^2$Lawrence Berkeley National Laboratory, Berkeley, CA, 94720, USA}

\begin{abstract}
We present the high-energy neutrino Monte Carlo event generator
ANIS (All Neutrino Interaction Simulation). The program provides a
detailed and flexible neutrino event simulation for high-energy
neutrino detectors, such as AMANDA, ANTARES or ICECUBE. It
generates neutrinos of any flavor according to a specified flux
and propagates them through the Earth. In a final step neutrino
interactions are simulated within a specified volume. All relevant
standard model processes are implemented. We discuss strengths and
limitations of the program.
\bigskip 

\noindent {\bf Program Summary} \newline \noindent {\it Title of
program:} ANIS
\newline {\it Program is obtainable from:} \aniswww
\newline {\it Computer on which the program has been thoroughly
tested:} Intel-Pentium based Personal Computers\newline {\it
Operating system:} Linux
\newline {\it Programming language used:} C++
\newline {\it
Memory required to execute: } 13 megabyte
\newline {\it Number of lines in distributed program (version 1.8):} 3300
\newline {\it Libraries used by ANIS:} HepMC \cite{hepmc}, CLHEP
vector package \cite{clhep}
\newline {\it Nature of physical problem:} Monte Carlo neutrino
event  generator for high-energy neutrino telescopes \newline {\it
Method of solution:}  Neutrino events are first
sampled according a specified flux, then  propagated through the Earth and
finally are allowed to interact inside a detection
volume.
\newline {\it Restrictions of the program:} Neutrino energies
range from 10~GeV to $10^{12}$~GeV.
\newline {\it Typical running time:} $10^4$ events require
typically a 1-GHz CPU time of about 300~s
\end{abstract}

\end{frontmatter}
\vfill\eject

\section{Introduction}
The first generation of open water/ice Cherenkov neutrino
telescopes take data (AMANDA \cite{amanda}, BAIKAL \cite{baikal})
or are under construction (ANTARES \cite{antares}, NESTOR
\cite{nestor}) while the second generation projects such as
IceCube \cite{icecube} are already in the planning phase. Such
devices aim at detection of cosmic high-energy neutrinos.
Cherenkov photons from charged particles produced in neutrino
interactions near or inside  the detector are used to identify
neutrinos. Because of their large volume and coarse granularity,
the energy threshold of these {\it open} instruments is typically
between 10~GeV and 100~GeV and therefore high when compared to
that of closed underground Cherenkov detectors, such as
SuperKamiokande and SNO. The flux of neutrinos produced by
interactions of cosmic rays in the atmosphere (so called
atmospheric neutrinos) falls steeply with energy ($\frac{\d N}{\d
E_\nu}\propto E_\nu^{-3.7}$). Hence the event rate due to
atmospheric neutrinos is largest near the detection threshold
energy. This is different for astrophysical neutrino fluxes which
generally are assumed to have harder spectra. Potential sources
for astrophysical neutrinos could be astrophysical objects such as
Active Galactic Nuclei (AGN), Gamma Ray Bursts (GRBs) or heavy
relic particles, such as produced e.g.\ in decays of Topological
Defects (TD). The relevant range of neutrino energies from the
more conventional sources such as AGN and GRBs range from
$10^3$~GeV, to $10^9$~GeV, energies while the range of neutrinos
expected from TD models extends up to $\sim 10^{15}$~GeV energies.

Apart from the desired astrophysical neutrinos and
atmospheric neutrinos, high-energy neutrino telescopes detect muons
produced in cosmic ray interaction in the atmosphere (so called
atmospheric muons). Due to the
low interaction probability of neutrinos, the trigger rates due to
atmospheric muons is generally orders of magnitude higher than
that of atmospheric neutrinos. Hence a large number of search strategies are
used, for example using the Earth as a shield and selecting only events with
horizontal and upward going directions.

In the analysis of the data of neutrino telescopes, Monte Carlo
(MC) simulations play a decisive role in designing and testing
event reconstructions and filters, determining filter efficiencies
and interpreting the results. Various programs exist for event
simulations. For example, atmospheric muons can conveniently be
simulated with the air-shower program CORSIKA \cite{corsika}. The
generated muons are further propagated through the overburden
consisting of water or ice using muon propagation programs such as
MMC \cite{mmc}, MUM \cite{mum} or PROPMU \cite{propmu}. In a final
step the detector response needs to be simulated using detector
specific programs. Programs for simulating neutrino events are
known to exist within various collaborations, however, to our
knowledge they are not publicly available. In particular most
programs  simulate only some specific
classes of events (for example $\nu_\mu$ reactions) in a limited
range of energies.

Here, we present the program ANIS for simulating neutrino events
of all flavors in a wide range of energies. ANIS is a Monte Carlo
event generator, which prepares neutrinos,
propagates them through the Earth and in a last step simulates a
neutrino interaction within a specified volume around the
detector. The aim of the program is to provide a tool for precise
simulation of neutrino events of all flavors in the energy range
addressed by high-energy neutrino telescopes. The program has been
designed  to fulfill the requirements for
analysis of AMANDA data, and could hence be of use for other
neutrino telescopes.

An application of ANIS was demonstrated in \cite{anis_icrc}, where
the expected rates for $\nu_e$ events as calculated with ANIS were
presented. The source code as well as further documentation
can be obtained from the project web site: \aniswww .


The paper is organized as follows. First a general description of
the program is given in Section \ref{description}. The implemented
neutrino interaction processes are shortly discussed in Section
\ref{interactions}. The possibility to add a new (non-standard
model) interaction process to ANIS is demonstrated there by an
example. In Sections \ref{tau} and \ref{propagation} the role of
tau leptons  and how neutrino propagation
is simulated in ANIS are discussed.
A description of the treatment of the final neutrino interactions
inside the detection volume follows in Section \ref{volume}.
Additional information relevant for potential users is
given in the appendix.

\section{Description of the program} \label{description} %

ANIS is written in C++ and uses the vector package of the CLHEP
library \cite{clhep} as well as the HepMC Monte Carlo event record
\cite{hepmc}. The event record holds all interaction vertices with
their respective incoming and outgoing particles. The currently
implemented interaction channels include charged current (CC) and
neutral current (NC) interactions as well as resonant $W^-$ production: 
$\bar \nu_e e^- \rightarrow W^- \rightarrow anything$.

Primary neutrinos are randomly generated on the surface of the
Earth. Currently only power-law  energy
spectra $\phi_\nu (E) \propto E^{-\alpha}$ are
implemented, however the code is flexible enough to
allow its extension to arbitrary neutrino spectra.

ANIS propagates neutrinos in small steps towards the detector. In
interactions with matter they are either absorbed (CC-case) or
regenerated at lower energies (NC-case). In the special case of CC
$\nu_\tau$ interaction, a short-living $\tau$-lepton is produced.
It propagates in matter, thereby loosing part of its energy, and
finally decays giving rise to a secondary $\nu_\tau$ and, in $\sim
17~\% $ of the  cases, to secondary $\nu_\mu$ or $\nu_e$.
The density profile of the Earth used for neutrino propagation is
chosen according to the Preliminary Earth Model \cite{prem}.

Once the detection volume is reached, a final vertex is generated
along the neutrino trajectory  within the detection volume. In the
case of a CC $\nu_\mu N$-interaction, ANIS correctly simulates the
muon scattering angle.

Along with the full event, four weights are stored: a
normalization constant,  a weight proportional to the total
interaction probability of the neutrino as well as two  weights corresponding 
to the atmospheric flux of electron and muon neutrinos \cite{lipari}.

Event rates for atmospheric and various extraterrestrial neutrino
spectra  are obtained by applying the appropriate weights to the
events. This last  step can conveniently be done by a user defined
energy dependent weight function applied within  analysis programs
as e.g.\ PAW \cite{paw} and  ROOT \cite{root}. Applications of the 
weights are further discussed in appendix \ref{sec:use}.

There are currently two output formats implemented: i) the AMANDA specific 
event format \emph{f2000} \cite{f2000} and ii) the HepMC \emph{ascii} format \cite{hepmc}.
\section{Neutrino cross sections} \label{interactions}%

At $E_\nu \lesssim 1 \times 10^6$~GeV, deep inelastic $\nu
N$-\crss\ may be successfully described in the framework of pQCD.
Parameterization of the $\nu N$\ \stfs, $F_i^{\nu N}(x,Q^2)$, may
be chosen e.g.\ according to CTEQ5 \cite{cteq}. Here, $-Q^2=q^2$
is the invariant momentum transfer, with the 4-momentum $q = k_\nu
-p_l$ given by the difference in momenta between the incoming
neutrino and outgoing lepton. The "Bjorken" variable $x$ is given
by $x=Q^2/(2M_N(E_\nu-E_l))$ where $E_\nu$ and $E_l$ are the
energies of the incoming neutrino and outgoing lepton.

At higher energies  cross-sections are dominated by scattering off
sea-quarks with small $x$. The unknown behavior of the \stfs\ at
$x \lsim 10^{-6}$ makes the calculations model dependent. The
uncertainty in extrapolations of $F_i^{\nu N}(x,Q^2)$ to small $x$
and large $Q^2$ influences the expected cross-sections at high
energies. Two options are presently included in ANIS: i) the
smooth both with respect to $x$ and  to $Q^2$ power-law
extrapolation of the pQCD CTEQ5 parameterization to small $x$ and
large $Q^2$, and ii) hard pomeron \cite{pom} enhanced
extrapolation \cite{hp}. The \crss\ of the first case, denoted in
the Fig.~1a as pQCD, practically coincide with \cite{gandhi},
while the second model (HP, dash-dotted curves) predicts
approximately $2$ times higher \crss\ at $E_\nu =1 \times
10^{12}$~GeV.

A kinematically important variable is $y=1-\frac{E_l}{E_\nu}$,
which characterizes the energy transfer to the outgoing lepton and
to the final hadronic state. The $y$-distributions, normalized to
the corresponding \crss, are plotted in Fig.~1b; the solid and
dashed curves stand for CC and NC $\nu N$-interactions,
respectively. The scattering angle, $\theta$, between incoming
neutrino and outgoing lepton is given by
\[ %
\cos\theta = 1 - \frac{x y}{1-y} \cdot \frac{m_{\rm N}}{E_{\nu}}.
\] %
The used cross-section data for CC- and NC-reactions are taken
from pre-calculated tables. The total cross-section is obtained
through interpolation between energy bins. A final state can be
represented by the pair of $x$ and $y$. Large sets of $x$ and $y$
have been generated for different energies and are stored in data
tables. During generation of neutrino events such pairs are
randomly drawn from the tables. The use of pre-calculated tables
makes the program fast and independent of other packages.

\begin{figure}[t]
\begin{minipage}[b]{.45\linewidth}
  \centering
    \includegraphics[height=15pc]{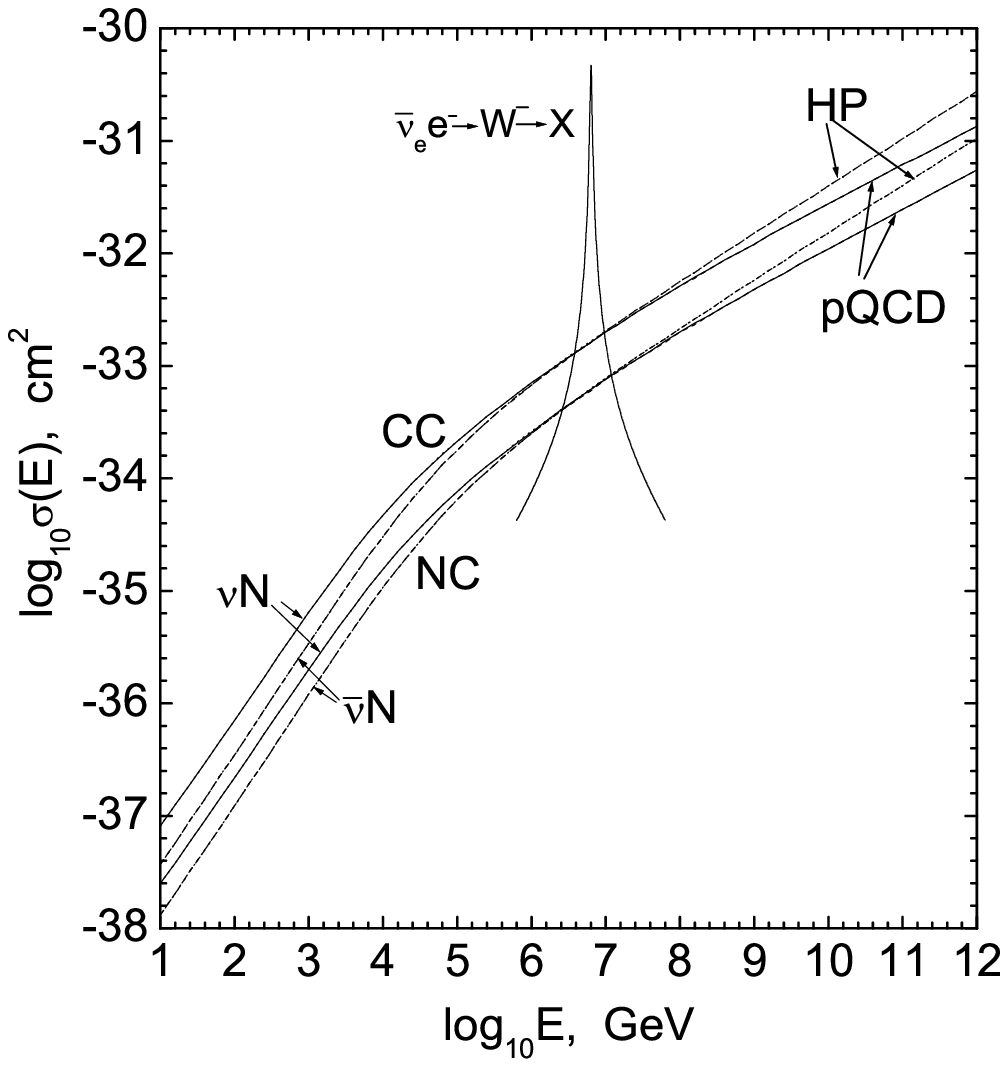}%
\end{minipage}\hfill
\begin{minipage}[b]{.45\linewidth}
  \centering
    \includegraphics[height=15pc]{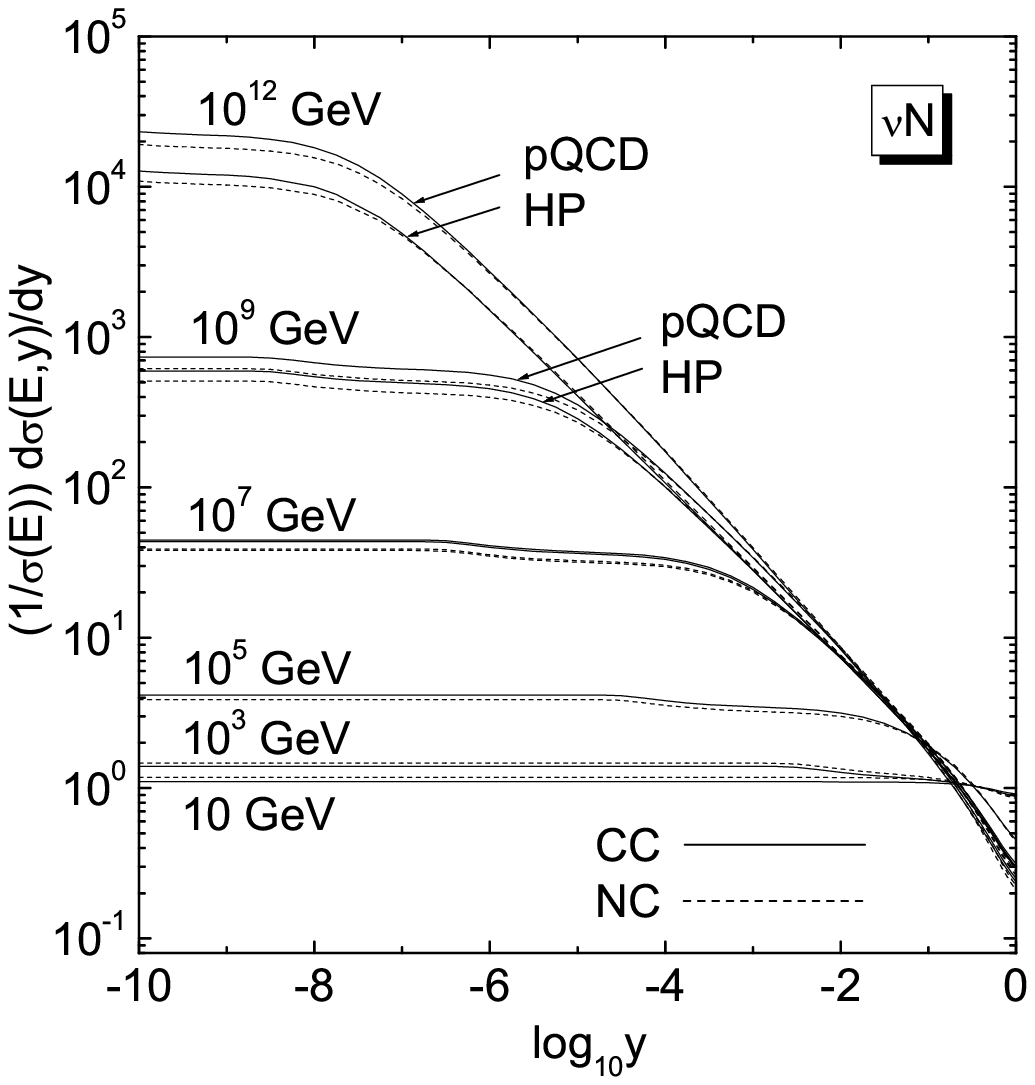}%
\end{minipage}

  \vspace{-0.5pc}
  \caption{a) \nuN- and resonance $\bar \nu_e e^- $-\crss\  and b)
  normalized $y$-distributions.}
\end{figure}

Since $m_e \ll m_N$,  at high energies practically all $\nu
e^-$-\crss\ are negligibly small, with the exception of resonant
$W^-$ production $\bar \nu_e e^- \rightarrow W^- \rightarrow
anything$ at $E_{\bar \nu_e} \approx 6.3\cdot10^6$~GeV (the so
called Glashow resonance, \cite{glashow}) (see Fig.~1a).
This  process is also included in ANIS.

ANIS was designed such that new cross-sections and processes can
be added  at a later stage.
Assuming for example, one would like to allow microscopic black
hole production \cite{bh} one has to derive a new (C++)
class for this process and provide the relevant member functions:
\newline

\noindent {\tt \small
  class SigmaBH : public Sigma \{\\
$~~$ public: \\
$~~~~$ double GetSigma(const GenParticle* neutrino, double material[]); \\
$~~~~$   void   FillVertex(GenVertex* vertex, double material[]); \\
\}\\
}
\\
{\tt GetSigma} returns the cross section, and the function {\tt
FillVertex} inserts the final state. By adding this new process to
the list
  of know processes, it will be included during simulation.
Thereby all particles  produced in the final states are written to
the event record and in case of neutrinos are further propagated.
In this way, any potential implications of non-standard model
physics can easily be tested.

Already implemented SM processes can be modified even without
modification of the code, simply by replacement of the
cross-section data tables obtained using other programs.

\section{Treatment of tau leptons} \label{tau}

Tau leptons can be  produced in CC $\nu_\tau N$-reactions and in
resonance  $\bar \nu_e e^- \rightarrow W^- \rightarrow \bar
\nu_\tau \tau^-$ scattering. The average tau decay length is
$l_\tau=49~{\rm m} \times (E_\tau / {\rm PeV})$. For energies
larger than $E_\tau \sim10^8 \rm GeV$, the energy loss during propagation 
has to be taken into account, leading to a slower growth of the decay
length. In the current version of the program, the stochastic
effects in the energy loss rate are neglected. The
energy loss rate is assumed to be continuous:
\begin{equation}
\frac{\d E_\tau}{\d X}=\alpha+ \beta(E) E_\tau,
\end{equation}
with $\beta=(1.508 + 6.3 \cdot (E/10^9~{\rm GeV})^{0.2}) \cdot
10^{-7} {\rm
      cm^2/g}$ \cite{range}.

In contrast to muon propagation, neglecting stochastic energy
losses does not alter the average $\tau$ range
 significantly \cite{dima2}. At energies of $E_\tau \sim 10^8$~GeV, the
continuous case is essentially indistinguishable from the
stochastic case and at energies around $E_\tau  \sim
10^{11}$ GeV the difference between the two cases is only
a few percent.  The reason, next to the energetic tau decay, is
that in case of taus (hard) bremsstrahlung is suppressed while
(softer) nuclear interactions dominate.

In ANIS, the tau is propagated in small steps of energy. Once the
age in the tau rest-frame exceeds the previously determined
lifetime the tau decay routines are called.

The tau decay is simulated with the help of TAUOLA \cite{tauola}.
Instead of linking directly to TAUOLA, we have created tables of
final states, consisting of the fractional energy: hadronic
($\pi^{\pm},K,..$), electro-magnetic (e, $\pi^0$ and $\gamma$),
muons, and the various neutrino flavors. Thereby we have assumed
left-handed polarization of the tau leptons, which is a good
approximation for neutrino interactions, if $E_\tau/m_\tau\gg1$.
The tau decay products are added to the event record and in case
of neutrinos are further propagated.

\section{Neutrino Propagation} \label{propagation}%

Neutrinos are generated at the surface of the Earth and then
propagated to the detector. They are 
assumed to travel along straight trajectories, neglecting the small
angle neutrino scattering. This is a good approximation, since at
energies below $10^5$~GeV when the average deflection angle (for
example through NC reactions) becomes larger $0.1^\circ$, the
neutrino interaction probability in the Earth is small. A
deflection angle of $0.1^\circ$  is below the current instrumental
resolution of existing or planned neutrino telescopes. The density
profile of the Earth is chosen according to the   Preliminary
Earth Model \cite{prem}. Neutrinos are propagated in small steps
through the Earth and in case of an interaction one of the active
interaction channels is chosen according the inclusive
cross-section. Its final state member function is invoked, which
adds the final state of the interaction to the event record.
Naturally, all propagation effects such as neutrino absorption,
$\nu_\tau \rightarrow \tau \rightarrow \nu_\tau$ regeneration,
regeneration in NC interactions as well as possible neutrino
production in the resonant scattering of electrons (e.g., $ \bar \nu_e
\rightarrow W^- \rightarrow \nu_{e,\mu,\tau}$) are included.

\section{Detection volume and final vertex} \label{volume}%

The size of the detection volume needs to be specified in the
steering file. The shape of the volume is a cylinder with its
$z$-axis parallel to the neutrino directions. Since open water/ice
Cherenkov detectors have a sensitivity to events with vertices
outside their fiducial volumes, the size of the cylinder should be
chosen appropriately. In particular due to the potentially long
range of muon and tau leptons, the final cylinder should extend to
the direction of the neutrino origin. As an alternative, one can
choose a cylinder with a variable length which adjusts to
the energy-dependent maximally possible muon and tau range. The
latter allows a more efficient simulation.

\begin{figure}[ht]
  \centering
    \includegraphics[height=15pc]{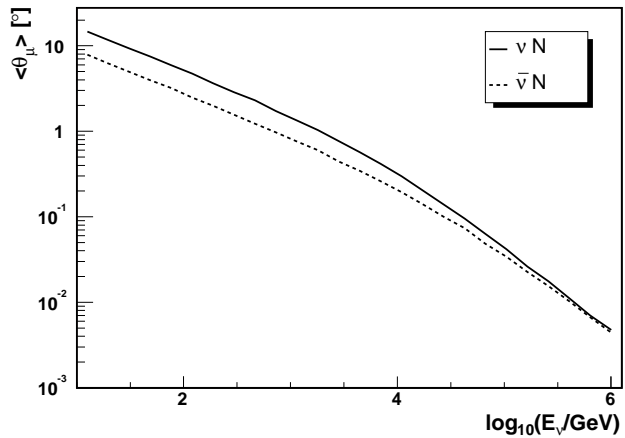}%
  \vspace{-0.5pc}
  \caption{Average angle $\theta$ between the incoming
  neutrino and the outgoing muon in a CC reaction as a
  function of the neutrino energy.}
\label{theta}

\end{figure}

Neutrinos reaching the detection volume are forced to interact
within this volume. The interaction probability is stored in the
event record. Two options are implemented: i)  all events are
written to disk -- in order to obtain a physical
spectrum, one needs to weight all events with their interaction
probability, and ii) events are sampled according their
interaction probability -- the events are sampled
using an acceptance-rejection method, where the selection criteria
are determined at run time during initialization.

The final state depends on the flavor of the interacting neutrino,
the type of the reaction as well as on the variables $x$ and $y$
of the interaction. Since \cher neutrino telescopes cannot
distinguish the different types of hadrons, the hadronization of
the final state is omitted and all hadronic energy is written out
as pions. Since neutrino telescopes have a muon angular resolution
of $\mathcal{O}(1^\circ$), the scattering angle between incoming
neutrino and outgoing muon needs to be taken into
account. Fig. \ref{theta} shows the average scattering angle for
neutrinos and anti-neutrinos as obtained with ANIS.

All muons of the final state have to be further propagated using
dedicated muon propagations programs (such as MMC or MUM), before
the events are passed to the detector simulation.

\section{Conclusion}
A new program for the simulation of neutrino induced events for
high-energy neutrino telescopes  is
presented. The program fulfills current requirements for data
analysis of neutrino telescopes, and is designed flexible such
that new processes could be added in the future.
The current version of ANIS includes all relevant SM
neutrino interaction processes. The energies for which
cross-section data is provided ranges from 10~GeV to
$10^{12}$~GeV.

\newpage
\centerline{\bf{Acknowledgments}} We acknowledge the contributions of Tonio 
Hauschild, suggestions and feedback from Stephan Hundertmark, as well as
discussions with Gary Hill.

\newpage

\begin{appendix}

\section{Organization of the Program}
%

The ANIS source code consists of several files in two directories:
anis/ and sigma/. The first directory contains the part of the
code    related to the simulation of events, while the latter
contains the part of the code related to interaction and decay of
particles.

{\bf anis/} \newline { anis.cc } -- the main file of the program
which calls the main processes. \newline { Steer.cc } -- the
steering file format. \newline { AnisEvent.cc} -- the ANIS event
record definition. \newline { FluxDriver.cc } -- contains classes
for simulating neutrino fluxes. \newline { NeutrTraject.cc } --
contains classes for calculating neutrino trajectories. \newline {
Propagation.cc }  --  contains classes related to the propagation
of neutrinos through the Earth to the detector. \newline {
FinalVolume.cc } -- contains classes related to the final
detection volume. \newline {\ FinalVertex.cc } -- classes for
sampling the vertex within the final volume. \newline { Earth.cc}
-- contains information on the Earth density profile.     \newline
{ AnisIO.cc} --  definition of various output streams. (see also
AnisEvent.cc).

{ \bf sigma/}\newline { Sigma.cc} -- the virtual base class for
all interaction processes.\newline { SigmaSM.cc} -- all Standard
Model processes.  \newline { SigmaTotal.cc} -- the container class
holding all activated processes. \newline { ProcessManager.cc} --
controls the active processes for the
 neutrino flavors and target materials.   \newline
{ Decay.cc} -- contains all decay classes. \newline { Table.cc} --
contains various utility classes for reading the data tables.
\newline

Next we describe the most relevant classes and member functions:

{ \tt  class FluxDriver \\
void Inject (AnisEvent *evt)} \\
 - This method inserts the initial neutrino into the event record.
The position of the neutrino origin is chosen on the Earth surface,
such that the straight trajectory of the neutrino intersects the final
detection volume.

{\tt  class Propagate\\
 void ThroughEarth (AnisEvent *evt)} \\
- this method propagates all neutrinos in the event record up to the 
backside of the the final detection volume.

{\tt class SigmaTotal\\
double  GetSigma (HepMC::GenParticle *nuin, double material[])\\
void    FillVertex (HepMC::GenVertex *vtx, double material[]) \\ }
- The {\tt GetSigma} member function returns the total cross-section for a
given neutrino and material (in the current version of the code isospin
invariance is assumed:  all interaction material contains equal numbers
of neutrons, protons and electrons).
The {\tt FillVertex} member function samples an interaction type from the
list of possible processes and fills the vertex record with corresponding
outgoing particles.

{\tt class ParticleDecay\\
void AddVertex (HepMC::GenParticle *parent, AnisEvent *evt)\\}
- this method inserts a decay vertex for the incoming {\it parent} particle
into the event record. So far the only unstable particle implemented
is the tau-lepton.

{\tt  class FinalVertex\\
void    ThrowVertex (AnisEvent *evt)}\\
- this method inserts a final neutrino interaction (vertex) located
within the final detection volume into the event record. The interacting
neutrino is chosen (with a probability proportional to the its interaction
probability) from all neutrinos reaching the final detection volume.

Further there is the control class {\tt Steer}, which holds the
run parameter of ANIS as well as the {\tt ProcessManager} which
controls the list of available interaction processes.

Further documentation of the code can be found on the
project web site: \aniswww.

\section{The steering file}
The program ANIS reads a steering file called "steer.dat". In case
a file name is given in the command line, this file is taken as
the steering file. The steering file contains all run specific
information.
Here is an example:\\
\vspace{0.1cm}

\noindent {\tt \small
\# The flux information \\
f p 1 1e2 1e4 -1. 1. 12 \\
\# The geometry information (volume is cylinder) \\
g c 300. 300. 300. 1.7e3 1 \\
\# The run information \\
r 100 1 1 2123421  \\
\# The cross-section processes and data.  \\
s CC cteq5/cteq5\_cc\_nu.data cteq5/final\_cteq5\_cc\_nu.data 101010 110 \\
s CC cteq5/cteq5\_cc\_nubar.data cteq5/final\_cteq5\_cc\_nubar.data 10101 110 \\
s NC cteq5/cteq5\_nc\_nu.data cteq5/final\_cteq5\_nc\_nu.data 101010 110 \\
s NC cteq5/cteq5\_nc\_nubar.data cteq5/final\_cteq5\_nc\_nubar.data 10101 110 \\
s GR dummy dummy 10000 1 \\
\# the data directory \\
d ../data } \vspace{0.3cm}

\noindent

Lines starting with a letter: s,f,g,r,d specify a program command
line. Lines starting with \# are for comments only.
\begin{itemize}
\item[f] (flux):
The f must be followed by a parameter specifying the spectrum type
(p for power law, this is so far the only implemented one), the
spectral index, the minimal neutrino energy, the maximal neutrino
energy (in GeV), the minimal cosine of the zenith angle of the
neutrino direction, and the maximal allowed cosine of the zenith
angle. The last number indicates the neutrino type with the
following possibilities: $12(=\nu_e+\bar{\nu}_e)$,
$14(=\nu_\mu+\bar{\nu}_\mu$), $16(=\nu_\tau+\bar{\nu}_\tau$).

\item[g] (geometry):
The g is followed by the type of the final volume (c stands for
cylinder, which is the only one implemented so far), the radius
(in meter) and the +/- height of the volume. The height is given
by two numbers, the first is the positive height (target region),
and the second is the height in the detector region. The next
element is the depth of the detector from the surface (in meter).
The final number is optional: if set to 1,  a geometry
optimization is invoked: if the lepton range is larger than the
specified volume, the volume is adjusted to the maximal lepton
range. So far this option is implemented only for muon-neutrino. 
If the number is missing or set to 0, then no re-adjustment is made.

\item[r] (run parameter):
The r is followed by the number of events to be simulated  and the
output format (1=HepMC,2=f2000). The third value indicates if
weighted (1) or unweighted (0) events should be simulated. The
last number is an optional seed for the generator. If no seed is
present, the system time is used to seed the random number
generator.

\item[s] (sigma):
The information following the s is the type of the process (CC:
charged current, NC: neutral current, GR: Glashow resonance) the
file containing the cross-section data, the file containing the
final state data, the active neutrino bit mask. ($\nu_e
\bar{\nu}_e \nu_\mu \bar{\nu}_\mu \nu_\tau \bar{\nu}_\tau$, 101010
means all neutrinos, 10101 means all anti-neutrinos and 10000 means
only electron anti neutrinos), and the active target material bit
mask (110 means protons and neutrons, 1 means electrons).

\item[d] (data directory)
this is the path to the data directory, containing the cross
section, tau decay and neutrino flux data.
\end{itemize}

\begin{figure}[t]
  \centering
    \includegraphics[width=0.8\linewidth]{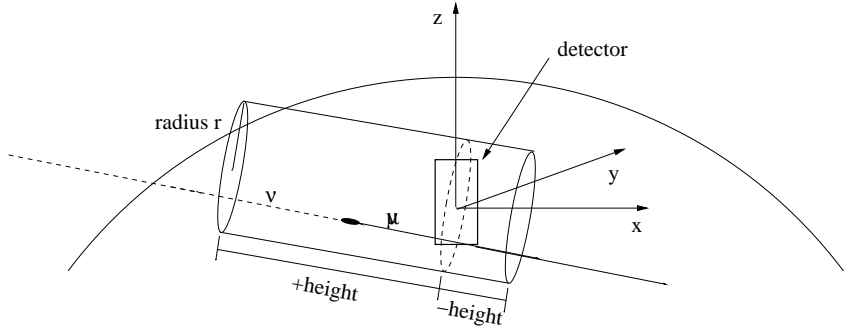}%
  \caption{The geometry: the coordinate system is anchored at the
  detector center, with the z-axis pointing away from the Earth center. The
 cylinder represents the {\it final  volume}, in which potentially detectable
neutrino interactions are simulated. Its size is adjustable through the
steering file.}

\label{geo}

\end{figure}

\section{Output}
Here only the HepMC event output format is discussed. An example
for a neutrino event as generated with ANIS is shown below. It is the first 
of 1000 events generated with the steering file described above.
\newcommand{\un}{\_\_\_\_\_\_\_\_\_\_\_\_\_\_\_}

\vspace{0.5cm} {\tt \small \noindent
GenEvent: \#1 ID=1 SignalProcessGenVertex Barcode: 0\\
 Current Memory Usage: 1 events, 2 vertices, 3 particles.\\
 Entries this event: 2 vertices, 3 particles.\\
 RndmState(0)=\\
 Wgts(4)=1.20431e-06 5.16006e+17 4.0188e-11 1.21211e-09\\
 EventScale -1 GeV       alphaQCD=-1     alphaQED=-1\\
                                    GenParticle Legend\\
        Barcode   PDG ID      ( Px,       Py,       Pz,     E ) Stat  DecayVtx\\
$\un \un \un \un \un \un$ \\
VVertex:       -1 ID:    1 (X,cT)=+5.37e+06,+1.26e+06,-2.98e+06,+0.00e+00 \\
 Wgts(1)=0 \\
 O: 1     10001       12 -4.32e+03,-1.01e+03,+2.40e+03,+5.05e+03   1        -2 \\
Vertex:       -2 ID:    0 (X,cT)=-3.50e+02,-7.01e+01,-8.75e+01,+0.00e+00 \\
 I: 1     10001       12 -4.32e+03,-1.01e+03,+2.40e+03,+5.05e+03   1        -2 \\
 O: 2     10002       11 -1.28e+03,-3.00e+02,+7.09e+02,+1.49e+03   1     (nil) \\
          10003      211 -3.05e+03,-7.15e+02,+1.69e+03,+3.56e+03   1     (nil) \\
$\un \un \un \un \un \un$ \\
}

Events consist of a list of vertices with attached incoming and
outgoing particles as well as a list of weight. For a description
of how to access events, vertex, particles and weights we refer to
the  HepMC documentation \cite{hepmc}.

\begin{itemize}
\item All vertex
positions are relative to the detector center (in units of
meters). The orientation is such that $z$ axis is pointing
upwards.
\item The particle identification numbers follow the Particle Data Group
scheme. Energy and momentum are given in units of
GeV.
\item Events have four weights attached: the interaction probability in the final volume ($P_{\rm int}$),
a normalization constant ($N$) and two weights  ($R_{\rm atm}^{\nu_e}$ and $R_{\rm atm}^{\nu_\mu}$)  corresponding to the
atmospheric neutrino flux for $\nu_e$ and $\nu_\mu$ \cite{lipari}.
\end{itemize}
\label{sec:weight}
\section{ Using ANIS weights}
\label{sec:use}
To obtain physical distributions of variables 
(such as neutrino energy or flavor at the detector) from ANIS events, one can 
conveniently use the weights described in appendix \ref{sec:weight}.

If events are generated according a spectral index 
$\gamma$ the application of an event weight  $N\cdot P_{\rm int}$ 
will result in a simulated flux \[\Phi(E)=(E/{\rm
GeV})^{-\gamma} [\rm GeV^{-1} s^{-1} sr^{-1} cm^{-2}]. \] 
The normalization is chosen such that 
$N\cdot \sum\limits_{i=1}^{N_{\rm event}}P_{{\rm int},i}$ 
results in the total  number of events  per year in the specified volume. 
Other spectra can be obtained by applying additional weight factors.
(For example, to obtain a flux with spectral index $\gamma^\prime$, the 
additional multiplicative weight should be $E^{\gamma - \gamma^\prime}$).

In case one is interested
in an atmospheric neutrino spectrum, each event should be weighted
according a weight  $N\cdot P_{\rm int}\cdot  R_{\rm atm}$, where the 
flux weight $R_{\rm atm}$ (which is provided by ANIS) should be
chosen according to the simulated neutrino flavor. 
Again, the normalization of the resulting distribution corresponds to
events per year in the specified volume. 

Accounting neutrino oscillations of atmospheric neutrinos can be done using weighting functions.
Assuming one is interested in muon neutrinos, and $P_{\nu_e \rightarrow \nu_\mu}$ is the energy and 
direction dependent probability for electron neutrinos to appear at the detector site as muon neutrinos, 
while   $P_{\nu_\mu \rightarrow \nu_\mu}$ is the probability of the initial 
muon neutrino to be detected as muon neutrino. The corresponding weight for each event is then:
 $N\cdot P_{\rm int}\cdot ( P_{\ \nu_\mu \rightarrow \nu_\mu} \cdot R_{\rm atm}^{\nu_\mu} +
P_{\ \nu_e \rightarrow \nu_\mu} \cdot R_{\rm atm}^{\nu_e})  $.

\end{appendix}

\end{document}